\documentclass[nofootinbib,aps,twocolumn]{revtex4}
\usepackage{amsmath,amssymb}
\usepackage{dcolumn}
\usepackage{bm}
\usepackage{xcolor}
\usepackage{graphicx}
\usepackage{commath}

\begin{document}

\title{A fractal LTB model cannot explain Dark Energy}

\author{Erick Pastén}
\email{erick.contreras[at]postgrado.uv.cl}

\author{V\'ictor H. C\'ardenas}
\email{victor.cardenas[at]uv.cl}

\affiliation{Instituto de F\'{\i}sica y Astronom\'ia, Universidad de
Valpara\'iso, Gran Breta\~na 1111, Valpara\'iso, Chile}

\begin{abstract}

We revisited the problem of describing, on average, a fractal distribution of matter using a Lemaitre-Tolman-Bondi (LTB) solution. Here we study the fractal structure of our local universe having a fractal dimension and a scale transition, ensuring an homogeneous bang-time function. We test our model with the latest type Ia supernova data, the Pantheon compilation, and discuss problems and possible improvements for it, concluding that a fractal transition in LTB cosmology cannot be used to explain the effects of dark energy without requiring an inhomogeneous big-bang, but it can be useful to study structures at low scales.

\end{abstract}

%\pacs{98.80.Cq}

\maketitle

%%%%%%%%%%%%%%%%%%%%%%%%%%%%%%%%%%%%%%%%%%%%%%%%%%%%%%%%%%%%%

\section{Introduction}

Dark Energy is the usual explanation for the apparent acceleration implied by the type Ia supernova (SNIA) data \cite{Perlmutter99}, \cite{Riess98}. However, the suggestion for the existence of dark energy is ultimately based on the cosmological principle, that of assuming global homogeneity and isotropy. The requirement of an extra parameter $\Omega_{\Lambda}$ is then necessary to explain the dimming of supernovae magnitude at large redshift.
%ok

Although at large scales the universe appears being homogeneous and isotropic in agreement to the CMB observations, at small scales it is far from being like that due to the presence of complex structures that produce underdensities \cite{Keenan13}, fractal-like structures \cite{Labini11},\cite{Labini98} and bulk flows that are not at rest with respect to the Hubble flow \cite{Feindt13} \cite{Hudson99} \cite{Magoulas16}. There have been many works claiming that some of these effects can mimic an apparent acceleration. Possibly the combination of many (even all) of these contributions may have a strong effect than we previously thought \cite{Celerier06} \cite{Enqvist07} \cite{Cosmai19} \cite{Tsagas11} \cite{Asvesta22}. As a step towards the understanding of the influence of the local structure in cosmology, in this work we explore a simple approach by studying the local fractal structure of the universe with the luminosity distance relations from SNIA data.
%ok

In a recent paper \cite{Cosmai19,Cosmai2}, the authors stated that a fractal-like $M(r)$ function can fit the SNIA data using a non-dark energy LTB model written as:
\begin{equation}\label{eq1}
    M(r) \propto r^D,
\end{equation}
where $D$ is the fractal dimension. Although interesting, the analysis is based on very special assumptions that make the model an extremely exceptional case. 
Here, we propose an analysis based also on the LTB metric but using both a fractal dimension and also a length-scale that seems to work without any theoretical problems, ensuring also an homogeneous big bang.
%ok

\section{The LTB model}

Lemaitre \cite{Lemaitre33}, Tolman \cite{Tolman34} and Bondi \cite{Bondi47} were the first to studied isotropic and radial inhomogeneous universes, known as LTB models. Here we restrict ourselves to the study of pressureless matter universes. Assuming for generality the existence of a cosmological constant with equation of state $ \rho_\Lambda=-p_\Lambda$, the energy-momentum tensor is given by:
\begin{equation}\label{eq2}
    T_\mu^\nu=-\rho_M(r,t)\delta^\mu_0\delta^0_\nu-\rho_\Lambda\delta^{\mu}_\nu,
\end{equation}
in which the matter density $\rho_M$ varies in time and radially in space, meanwhile $\rho_\Lambda$ remains fixed. The metric of a LTB universe can be written as:
\begin{equation}\label{eq3}
ds^2 = -dt^2 + \frac{R'^2}{1+2E(r)}dr^2 + R^2d\Omega^2,
\end{equation}
where $R=R(r,t)$ is a generalization of the scale factor and $E(r)$ is a function associated with the total energy. Here a prime indicates a derivative respect to $r$ and a dot a derivative respect to time $t$. From Einstein's equation we get the evolution equation:
\begin{equation}\label{eq4}
    \frac{\dot{R}^2}{R^2}=\frac{2GM(r)}{R^3}+\frac{8\pi G \rho_\Lambda}{3} +\frac{2E}{R^2},
\end{equation}
where $M(r)$ is an integration constant (in time). 
%ok

Notice that by writing $R(r,t)=a(t)r$ where $a(t)$ is the scale factor and $2E(r)=-Kr^2$ with $K$ being the curvature constant factor, the metric reduces to the FLRW metric and (\ref{eq4}) reduces to the usual Friedmann equation. 
%ok

The $M(r)$ function can be interpreted as the mass contained in a sphere of radius $R=ar$ in the FLRW limit. The matter density is defined through:
\begin{eqnarray}\label{eq5}
\frac{M'}{R'R^2} = 4\pi \rho _M.
\end{eqnarray}
So the mass function $M(r)$ is defined as the integral of $\rho_M$ in the LTB space as:
\begin{equation}\label{eq6}
    M(r)=\int_0^r\rho _M (4 \pi R'R^2 dr)
\end{equation}
Defining the local Hubble rate by
\begin{equation}\label{eq7}
    H(r,t)=\frac{\dot{R}(r,t)}{R(r,t)},
\end{equation}
we can write (\ref{eq4}) as:
\begin{equation}\label{eq8}
    H^2= \frac{2GM(r)}{R^3}+\frac{8\pi G \rho_\Lambda}{3} +\frac{2E}{R^2}
\end{equation}
By defining the local matter density parameter $\Omega_m$ as
\begin{equation}\label{eq9}
    2GM(r)=H_0^2(r)\Omega_m(r)R_0^3(r),
\end{equation}
where $H_0(r)=H(r,t_0)$ is the present Hubble local rate, $R_0=R(r,t_0)$ being an initial condition that can be gauged, and for the curvature term
\begin{equation}\label{eq10}
    2E(r)=H_0^2(r)\Omega_k(r)R_0^2(r),
\end{equation}
then, the local Friedman equation takes the form
\begin{equation}\label{eq11}
H^2(r,t)=H_0^2(r)( \Omega_m(r) A^3 + (1-\Omega_m(r))A^2),
\end{equation}
where $A(r,t)=R_0/R$. This means that we can choose any $R_0(r)$ function and the equations remains the same. This freedom is usually fixed by taking $R_0(r)=r$ \cite{Enqvist07}. The Friedman equation (\ref{eq11}) defines completely the relation between the functions $H_0(r)$, $\Omega_m(r)$, $R(r,t)$ and $H(r,t)$.
%ok

%There is seems that there is a strong degeneration in this equations due to the large quantity of parameters in the free functions that destroy any potential of LTB to be a predictive model, as it was stated by \cite{Celerier99}. However, we can constrain the model in some ways to reduce the quantity of parameters. 

One of the most used constraint to LTB models is based on the time that has passed from the big bang until now \cite{GarciaBellido08}, known as the bang time $t_{BT}(r)$ defined by:
\begin{equation}\label{eq12}
\begin{split}
    H_0(r)(t-&t_{BT}(r))=\int_0^{R(r,t)/R_0(t)}\frac{dx}{\sqrt{\Omega_m/x+\Omega_k}} \\
    &=\frac{R(r,t)}{R_0(r)\sqrt{\Omega_k(r)}}\sqrt{1+\frac{R_0(r)\Omega_m(r)}{R(r,t)\Omega_k(r)}} \\
    &-\frac{\Omega_m(r)}{\sqrt{\Omega_k(r)^3}}\sinh^{-1}{\sqrt\frac{\Omega_m(r)R(r,t)}{\Omega_k(r)R_0(r)}}.
\end{split}
\end{equation}
%Where 
This function means that not all locations in the universe were created at the same time. If we set $t=t_0$ we can obtain the current age of the universe $t_{BB}(r)$:
\begin{equation}\label{eq13}
\begin{split}
    &H_0(r)(t_0-t_{BB}(r))=H_0(r)t_{u}(r)=\int_0^{1}\frac{dx}{\sqrt{\Omega_m/x+\Omega_k}} \\
    &=\frac{1}{\sqrt{\Omega_k(r)}}\sqrt{1+\frac{\Omega_m(r)}{\Omega_k(r)}} 
    -\frac{\Omega_m(r)}{\sqrt{\Omega_k(r)^3}}\sinh^{-1}{\sqrt\frac{\Omega_m(r)}{\Omega_k(r)}}, 
\end{split}
\end{equation}
where we have defined $t_{u}(r)$ as the local age of the universe. It is interesting to note that we need to fix two functions here, chosen among $\Omega_m(r)$, $H_0(r)$ and $t_{BT}(r)$, and also fix a gauge (usually $R_0(r,t)=r$) to define completely the scale factor function $R(r,t)$. We obtain a similar result if we use instead the functions $M(r)$ and $E(r)$:
\begin{equation}\label{eq14}
\begin{split}
    \sqrt{2}(t-&t_{BT}(r))=\int_0^{R(r,t)}\frac{dx}{\sqrt{2GM(r)/x+2E(r)}} \\
    &=\frac{R(r,t)}{\sqrt{E}}\sqrt{1+\frac{GM(r)}{R(r,t)E(r)}} \\
    &-\frac{GM(r)}{\sqrt{E(r)^3}}\sinh^{-1}{\sqrt\frac{GM(r)R(r,t)}{E(r)}} 
\end{split}
\end{equation}
In this case, it seems we need to set three functions, $t_{BT}(r)$, $M(r)$ and $E(r)$, to define $R(r,t)$, but in fact one of them is a pure gauge. Usually $M(r) \propto r^3$ is assumed as this gauge (as we mentioned in the introduction, in some fractal-like models it has been proposed to use Eq.(\ref{eq1}) as the gauge, with $D$ being the fractal dimension \cite{Cosmai19}), so we just need to define two functions as we will see.
%ok

Another way to define LTB model is thinking in a density contrast profile $\delta(r,t)$ defined as:

\begin{equation}\label{eq15}
\delta(r,t)=\frac{\rho(r,t)-\rho_0^\infty}{\rho_0^\infty},
\end{equation}
where $\rho_0^\infty=\rho(\infty,t_0)$ is the mean density constant very far outside the local inhomogeneity at the present time. This function is useful because this allows us to describe the mass profile using the data directly from the observational photometric local surveys. We have for $t=t_0$:
\begin{equation}\label{eq16}
\rho_0(r)=\rho_0^\infty(\delta_0(r)+1)
\end{equation}
Although the functions $\Omega_m(r)$, $M(r)$ and $E(r)$ have a clear meaning, it is difficult to relate them directly to the observed quantity $\delta_0(r)$. So instead of using the complexities of (\ref{eq13}), we can approximate a solution by assuming the universe being described by FLRW and define a integrated density function as:
\begin{equation}\label{eq17}
\rho_0^H(r) \propto \frac{M(r)}{r^3} \propto (\delta^H(r)+1),
\end{equation}
which is exactly true for an homogeneous universe. Using the gauge $R_0(r)=r$ we can write:
\begin{equation}\label{eq18}
\delta^H(r)+1 \propto \frac{M(r)}{r^3}  \propto \Omega_m(r)H_0(r)^2,
\end{equation}
where $\delta^H(r)$ is the integrated density contrast. Then, constraining $t_{BT}(r)$ and $\delta^H(r)$ we can define completely $R(r,t)$. It is important to note that this is not exactly the integrated density contrast, but is a useful approximation that allows us to define the universe in terms of density real parameters. 
%ok

\section{Luminosity Distance}

The luminosity distance in LTB models can be obtained from:
\begin{equation}\label{eq19}
    d_L = (1+z)^2 R(r,t),
\end{equation}
where $r=r(z)$ and $t=t(z)$ are defined by the geodesic equations.
\begin{eqnarray}\label{eq20-21}
\frac{dr}{dz} &=& \frac{\sqrt{1-k}}{(1+z)\dot{R'}} \\
\frac{dt}{dz} &=& -\frac{R'}{(1+z)\dot{R'}}
\end{eqnarray}
These equations are necessary to make contact with the observations. In our particular case, we use the Pantheon sample \cite{Scolnic18} that release the redshift $z$, distance modulus $\mu(z)$ and the error $\sigma_{\mu}(z)$ for $1048$ type Ia supernova that can be compared to the theoretical expectation
\begin{equation}\label{eq44}
    \mu(z) = 5\log_{10}\frac{d_L(z)}{10 \text{pc}}.
\end{equation}

\section{Fractals in LTB Model}

In this section we follow the treatment of Ref. \cite{Cosmai19,Cosmai2}. In that work, the authors argue that from the statistical analysis of the 3D galaxy distribution, the average conditional density can be described locally as
\begin{equation}\label{eq22}
    \langle n(r) \rangle  \sim r^{-\gamma}
\end{equation}
where $\gamma$ is a phenomenological parameter that can take values from $0.2$ to $0.9$ in a scale range from $1$ Mpc/$h$ until $100$ Mpc/$h$. Beyond that scale, a transition to homogeneity ($\gamma=0$) is expected whose behavior is not completely clear. Then for the structure of a real galaxy we can approximate the density of matter as:
\begin{equation}\label{eq23}
    \rho_M(r) \sim \langle n(r) \rangle
\end{equation}
a behaviour that can be described using a LTB model. In this case, the function $M(r)$ is related to this fractal density function due to equation (\ref{eq6}):
\begin{equation}\label{eq24}
    M(r)=\int_0^r \langle n(r) \rangle(4 \pi R'R^2 dr),
\end{equation}
where $M(r)$ can be understood as the matter inside a volume of radius $r$. In \cite{Cosmai19}, it was stated that a fractal-like $M(r)$ function can fit the SNIA data in a pressureless LTB model given by
\begin{equation}\label{eq25}
    M(r)=\Phi r^D,
\end{equation}
where $D=3-\gamma$ is the fractal dimension and $\Phi$ is the mass-scale. Let us study this model at some detail. 

First of all, we have to emphasize a subtlety in the notation. The authors of \cite{Cosmai19} perform the integration of the equation
 \begin{equation}\label{eqadd1}
    \dot{R}^2+2R\Ddot{R}+k=0,
 \end{equation}
assuming that at $t=0$, the $R$ function takes the value $R_0$ from which leads the definition:
\begin{equation}\label{eq26}
    R(r,0)=R_0(r).
\end{equation}
In the LTB literature usually $R_0(r)$ is the {\it present} scale factor (which is usually a gauge in LTB cosmology when $\Omega_m(r)$ is used instead of $M(r)$), while $R(r,0)=0$ is the initial scale factor in the homogeneous big bang itself (usually setted to $0$). This leads to a confusion with the common definitions in LTB cosmology literature and should be pointed out.

Second, following the previous redefinition for the $R_0$ function, an inhomogeneous negative bang time function $t_{BT}(r)<0$, that notably is dependent on the fractal dimension of matter $D$ \cite{Cosmai2}, is obtained :
\begin{equation}
    t_{BT}=-\frac{2}{3}\sqrt{\frac{R_0^3(r)}{2GM(r)}}.
\end{equation}
However, an inhomogeneous bang time is something physically undesirable if we are trying to understand the influence of the fractal distribution of matter in cosmology. A similar problem was pointed out by \cite{Kenworthy19} discussing the work made in \cite{Hoscheit18} about the influence of a large void in LTB cosmology, as an inhomogenous big bang results in different ages of the universe at difference places of it. This introduces a new fundamental inhomogeneity that requires new physics in the same sense than the addition of dark energy. If we perform the integration of the equation for a purely flat universe, with an homogeneous bang time function and the function $M(r)$ with the usual limits ($t\in[0,t],\>R\in[0,R(r,t)]$ we just get the EdS results:
\begin{equation}\label{eq27}
    R(r,t)= \left (\frac{9M(r)}{2}\right )^{1/3}t^{2/3}.
\end{equation}
We can see here that the election of the function $M(r)$, with the assumption of global flatness and homogeneous big-bang is just a gauge related to the scale of the radial coordinate. Also, the Hubble function can be written as
\begin{eqnarray}\label{eq28}
    H(t)=\frac{\dot{R}}{R}= \frac{2}{3t}, 
\end{eqnarray}
independently of the $M(r)$ function used. We have no divergences for $H_0$ at $r=0$, but also the fractal structure described with $M(r)$ cannot be used to explain any $H_0$ tension.

The question now emerges, can we use LTB models with $E(r)=0$ to describe fractal distribution of matter and assuming an homogeneous big bang? In the following we describe our proposal to do that.

\subsection{Sharp transition model}

First, we propose to define a function $M(r)$ that can describe both, the internal fractal statistical behaviour of the density and also a transition to a FLRW universe (a transition between fractality and non-fractality). The function $M(r)$ could be described as
\begin{align}\label{eq29-30}
    M_{in}(r)&=\Phi_{in}r^D,  &r<L \\
    M_{out}(r)&=\Phi_{out}r^3, &r>L,
\end{align}
where $M\sim r^3$ is the mass for a FLRW universe. At this point we have four free parameters in the model: the mass scale $\Phi_{in}$ and $\Phi_{out}$, the fractal dimension $D$ and the scale of the fractal structure $L$. If we demand for continuity at $r=L$ we can reduce to three free parameters as:
\begin{eqnarray}\label{eq31}
   \Phi_{in}L^D=\Phi_{out}L^3,
\end{eqnarray}
then the model becomes
\begin{align}\label{eq32-33}
    M_{in}(r)&=\Phi_{out}L^3\left (\frac{r}{L}\right)^D, &r<L \\
    M_{out}(r)&=\Phi_{out}r^3, &r>L.
\end{align}
Defining the function in this way, the effects between the internal fractal structure and the external EdS universe should be noticeable. A similar approach was developed in \cite{Ruffini17}, but the difference is that they fix the scale of transition about $r=2300$ Mpc meanwhile for us this is a free parameter to test.
%ok

\begin{figure}[ht]
\centering
\includegraphics[scale=0.35]{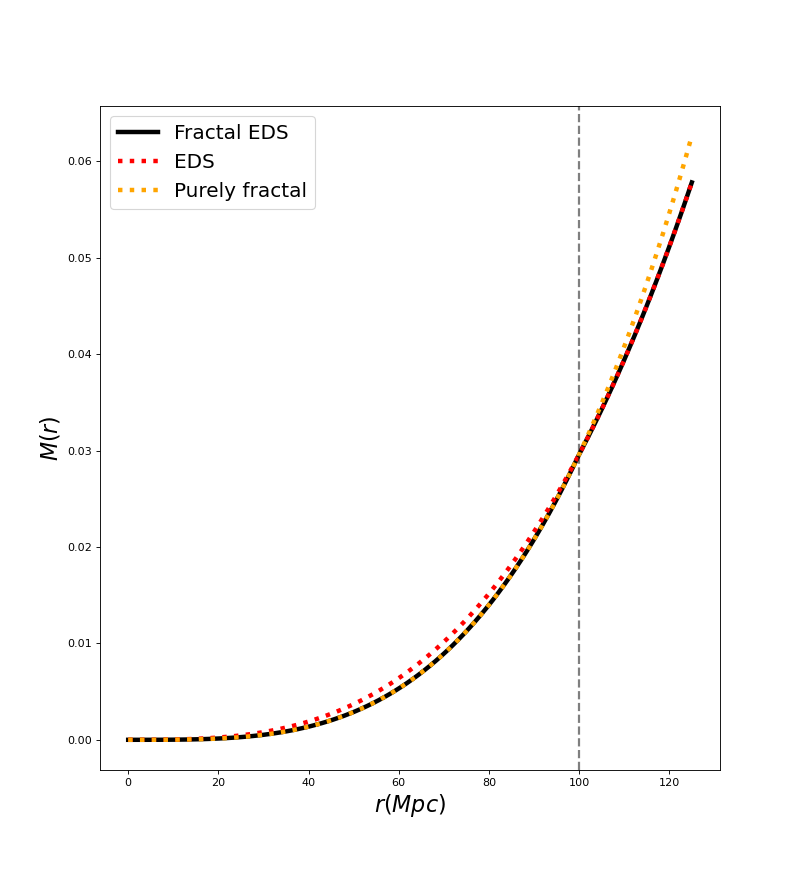}
\caption{$M(r)$ profiles. In black appears the fractal model with a sharp transition to EdS; in red pointed line EdS and in yellow pointed line a purely fractal universe. We use $D=3.36$, $L=100 Mpc$ and $h_{out} \sim 0.73$. The vertical gray line is the lenght $L$ of the transition. The functions $M_1(r)$ and $M_2(r)$ that describes a smooth transition in equations (\ref{add3}) and (\ref{add4}) are not plotted as those are very similar to the black line (see, e.g., fig. 2 of  \cite{Calcagni_2017}). }
\label{fig:1}
\end{figure}

\begin{figure}[ht]
\centering
\includegraphics[scale=0.35]{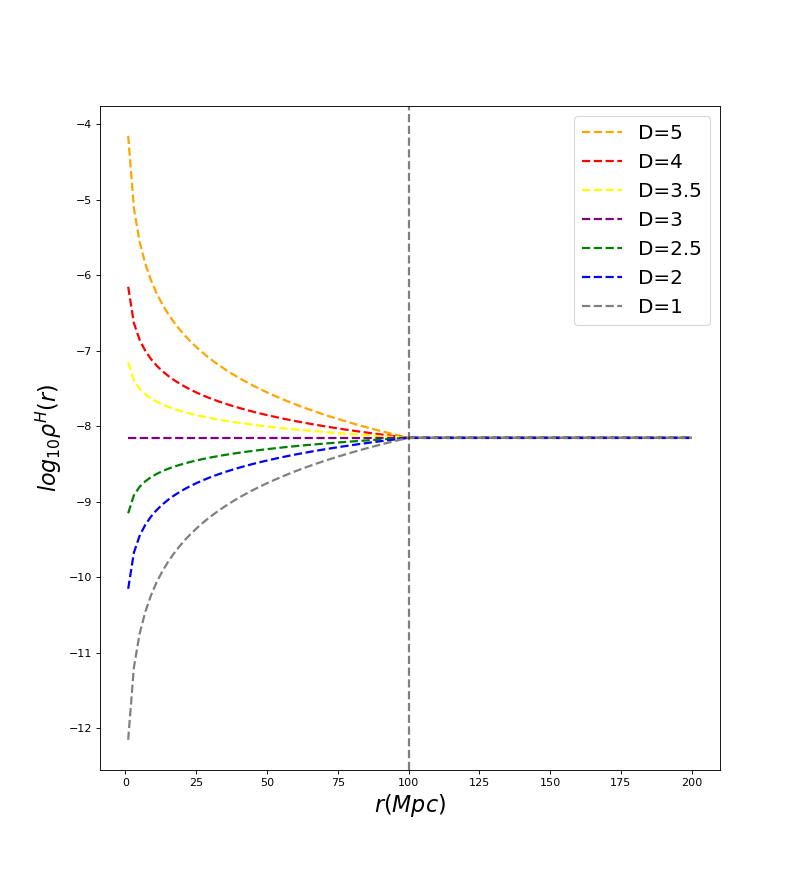}
\caption{Integrated density profiles for different values of $D$. We use $L=100 Mpc$ and $h_{out} \sim 0.73$. The vertical gray line is the length $L$ of the transition. Note that profiles with $D>3$ are similar to an over-dense region and $D<3$ to an under-dense region. $D=3$ correspond to the common constant integrated density profile in EdS universe. }
\label{fig:2}
\end{figure}

As a first approach, we assume $\Omega_{\Lambda}=0$ and the universe being just EdS beyond the transition. %This is justified as we use low redshift supernovaes and the effect of the Dark Energy is unnoticeable, but is debatable if we use the full sample of SNIA 
We use low redshift type Ia supernova data to estimate the transition scale for fractality. We also use the full sample \footnote{We can understand this approach as being a tool to estimate the fractal transition in low redshift data, but also as a possible explanation for dark energy with fractal structure using the full sample. However, this last approach is questionable as it does not exist a clear evidence for fractal structure for scales $r>100$ Mpc}, in which case we have to add another parameter to be fixed. Following \cite{Alexander09} and considering an EdS universe for $r>L$, we choose for $\Phi_{out}$ to be in agreement with the age of the universe in an EdS universe:
\begin{eqnarray}\label{eq34}
    \Phi_{out}=\frac{4 \pi}{3} \rho_0 = \frac{H_0^2}{2},
\end{eqnarray}
where $H_0 \sim h_{out}/3000$ Mpc is the Hubble parameter outside the fractal behaviour in the FLRW limit. With this scale we have the radial coordinate in Mpc and just two free parameters: $D$ and $L$. The $R(r,t)$ can then be written as:
\begin{eqnarray}\label{eq35-36}
    R_{in}(r,t) &=& \left(\frac{9\Phi_{in}}{2}\right)^{\frac{1}{3}}L\left (\frac{r}{L}\right)^{D/3}t^{2/3},\>\>r<L \\
    R_{out}(r,t) &=& \left(\frac{9\Phi_{out}}{2}\right)^{\frac{1}{3}} rt^{2/3},\>\>r>L, 
\end{eqnarray}
It is important to note that in this model, the density is surprisingly homogeneous. Using equation (\ref{eq11}) we can obtain the same density profile than an EdS universe:
\begin{eqnarray}\label{eq37}
     \rho_0(r)=\frac{3\Phi_{out}}{4\pi},
\end{eqnarray}
but the integrated density at $t_0$ is:
\begin{eqnarray}\label{eq38}
    \rho_0^H(r) \propto r^{D-3}=r^{-\gamma},\>\>r<L 
\end{eqnarray}
which is constant for $r>L$. Using the continuity condition at $L$
%\begin{eqnarray}\label{eq46}
%    \rho_0^H(L)=\frac{3M(L)}{4 \pi L^3}=\frac{3\Phi_{out}}{4\pi},
%\end{eqnarray}
we obtain:
\begin{equation}\label{eq39-40}
    \rho_0^H(r)= \left\{
  \begin{array}{lr} 
       \frac{3\Phi_{out}}{4\pi}\left(\frac{r}{L}\right)^{-\gamma} &, r<L \\
      \frac{3\Phi_{out}}{4\pi} &, r>L
      \end{array}
\right.
\end{equation}
%%%%%%%%%%%%%%%%%%%%%%%%%%%%%%
The behaviour of $M(r)$ and the densities are displayed in Fig. \ref{fig:1} and  Fig. \ref{fig:2}. It is interesting to note that the fractal behaviour can be also be defined as:
\begin{eqnarray}\label{eq41-42}
 \rho_0^H(r)&=&\rho_0 \left(\frac{r}{L}\right)^{-\gamma} ,\>\>r<L \\ 
 \rho_0^H(r)&=&\rho_0  ,\>\>r>L 
\end{eqnarray}

\begin{figure}[ht]
\centering
\includegraphics[scale=0.35]{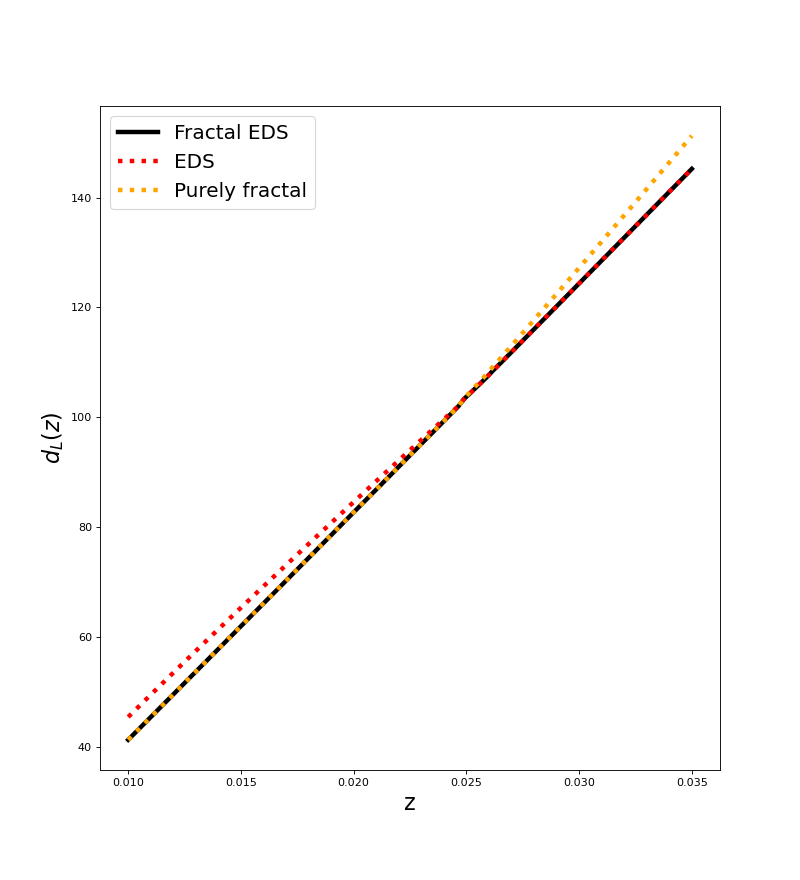}
\caption{Luminosity Distance for the different universes. Same description as in FIG. \ref{fig:1}. Note that we use the purely fractal universe with the scale $\Phi_{in}$ to illustrate.}
\label{fig:3}
\end{figure}

As simple it is, this model assumes a sharp transition which is not very realistic. Purely Fractal and Fractal-EdS universe are plotted in Fig. \ref{fig:3}.

\subsection{Smooth transition model}

A smooth slope transition around $r=L$ could be introduced considering a multi-power mass function. Motivated by the work developed in \cite{Calcagni_2017}, a binomial ``average" mass profile could be useful for this case. If $D<3$ we can write:

\begin{align}\label{add3}
    M_1(r)&=\Phi_{out}\left( L^3\left (\frac{r}{L}\right)^D + r^3 \right).
\end{align}
Note that as $D<3$, for $r \gg L$ the term $r^3$ dominates the mass function, meanwhile for $r \ll L$ the $r^D$ term is stronger. In the opposite case when $D>3$ we can propose:

\begin{align}\label{add4}
    M_2(r)&=\frac{\Phi_{out}}{\frac{1}{L^3}\left (\frac{L}{r}\right)^D+\frac{1}{r^3}},
\end{align}
which follows a similar asymptotic behaviour. We choose the same $\Phi_{out}$ value described in the last section to allow for consistent convergences at $r\gg L$. This approach has the advantage of being numerically friendlier as we have a unique mass profile for all $r$ space.

\section{Statistical analysis with SNIa data}

To test our models, we use the Pantheon sample comprised by 1048 SNIa in the redshift range $z \in (0.01, 2.3)$. As we are interested in studying a possible transition between fractal behaviour and a FLRW universe, we consider two approaches. First, using only low redshift SNIa, using a cutoff of $z<0.3$ corresponding approximately to $\sim 800 $Mpc. Second, using the full sample. We select $h_{out}\sim 0.73$ in the first case to match local Hubble parameter scale measures \cite{Dainotti_2021} and $h_{out}\sim 0.45$ in the second case, in order to converge for low scales of the Hubble parameter that fits CMB with an EdS universe \cite{Alnes}. The full Tripp Formulae for distance modulus is:

\begin{equation}\label{eq43}
    \mu_{obs}=m_b^*-M=m_b+\alpha x-\beta c+\Delta_M-M,
\end{equation}
here, $m_b$ corresponds to the peak apparent magnitude in the B-band, $M$ is the absolute B-band magnitude of a fiducial SNIa, $x$ and $c$ are light curve shape and color parameters, $\alpha$ is a coefficient of the relation between luminosity and stretch and $\beta$ is a coefficient of the relation between luminosity and colour, and $\Delta_M$ is a corrections based on the mass of the host galaxy. All of these parameters, usually called \textit{nuisance parameters}, are needed to standardize the SNIa data leading to the corrected magnitude $m_b^*$. Those correction are provided in Pantheon catalogue, and the parameter $M$ can be marginalized. This relation can be used to perform statistical analysis with the distance theoretical modulus for any cosmological model given by Eq.(\ref{eq44}).

\section{Results}

We use the code \texttt{EMCEE} \cite{2013PASP..125..306F} to test the models against the SNIa data. This is a pure python implementation of the affine invariant ensemble sampler for Markov chain Monte Carlo proposed by Goodman and Weare \cite{2010CAMCS...5...65G}. From this analysis we obtain the best fit values for $D$ and $L$ for each model. We used directly the corrected magnitudes from the catalogue, but also another approach allowing the nuisance parameters $\alpha$ and $\beta$ to vary along the cosmological parameters. In this last approach, we ignore the step mass correction as this does not impact significantly the cosmological fit.

\subsection{Sharp transition}

\begin{figure}[ht]
\centering
\includegraphics[scale=0.33]{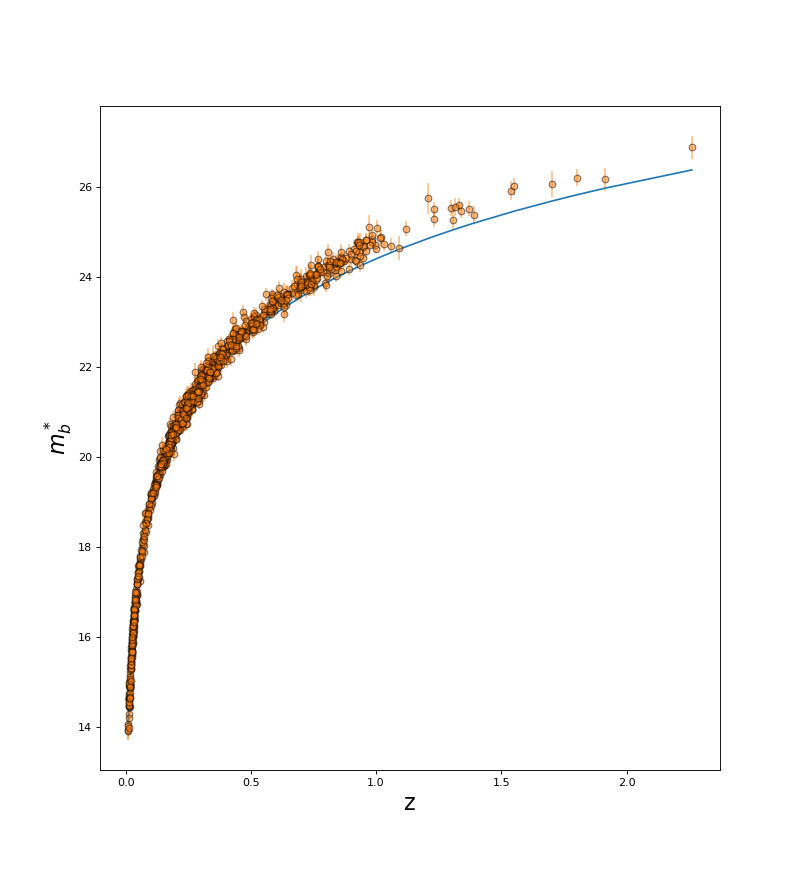}
\caption{Best fit function for corrected magnitudes of the full Pantheon sample using a fractal sharp transition. We get a notably worst fit than previous studies in fractal LTB models that use an inhomogeneous big bang.}
\label{fig:4}
\end{figure}

 The results for the sharp transition model and details are shown in Table \ref{Table:sharp} . We note that for the low redshift supernovae we can get a relatively good fit, but not for the full sample. We also note high errors in the parameter estimation when we do not use tight priors,possible attributed to the sharp transition at $r=L$. Best fit for the full sample is plotted in Figure \ref{fig:4}.
 
To look for insights into the value of the fractal dimension $D$, we also consider priors on the parameter $L$, which is our fractal transition scale. Instead of fixing it at a convenient value, as previous works have done, here we introduce Gaussian priors for $L$ using different values coming from large scale research of the transition scale for homogeneity. For example, in \cite{2010MNRAS.405.2009Y} the authors concluded that the scale of homogeneity should be less than $260h^{-1}$ Mpc. Also in \cite{2005MNRAS.364..601Y} and \cite{2005ApJ...624...54H} the authors have suggested even a smaller scale of $L=60h^{-1}$Mpc. Let us consider four values for $L$: $L=60h^{-1}$, $L=100h^{-1}$, $L=150h^{-1}$ and $L=200h^{-1}$. Considering Gaussian priors on this values with a 5\% of error, we find the results showed in the last panel in Table \ref{Table:sharp} for low redshift supernovae at $z<0.3$. Again, for sharp transition model the error on $D$ is very high. This means that a flat LTB fractal model with a sharp transition, homogeneous big bang and no cosmological constant does not perform well to fit SN IA data.

\begin{figure}[ht]
\centering
\includegraphics[scale=0.33]{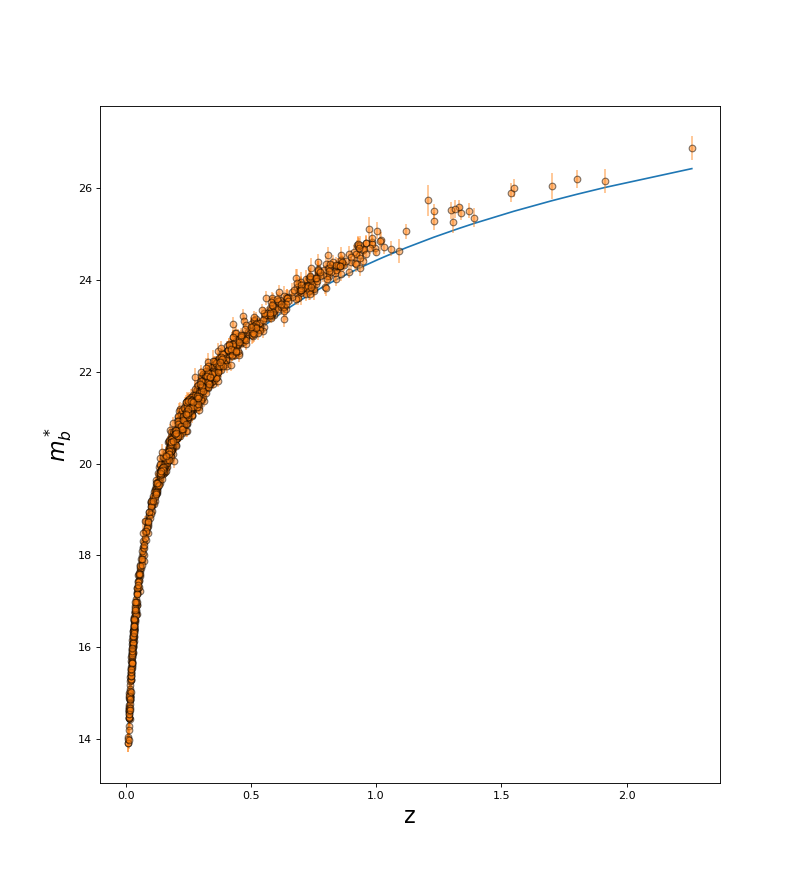}
\includegraphics[scale=0.33]{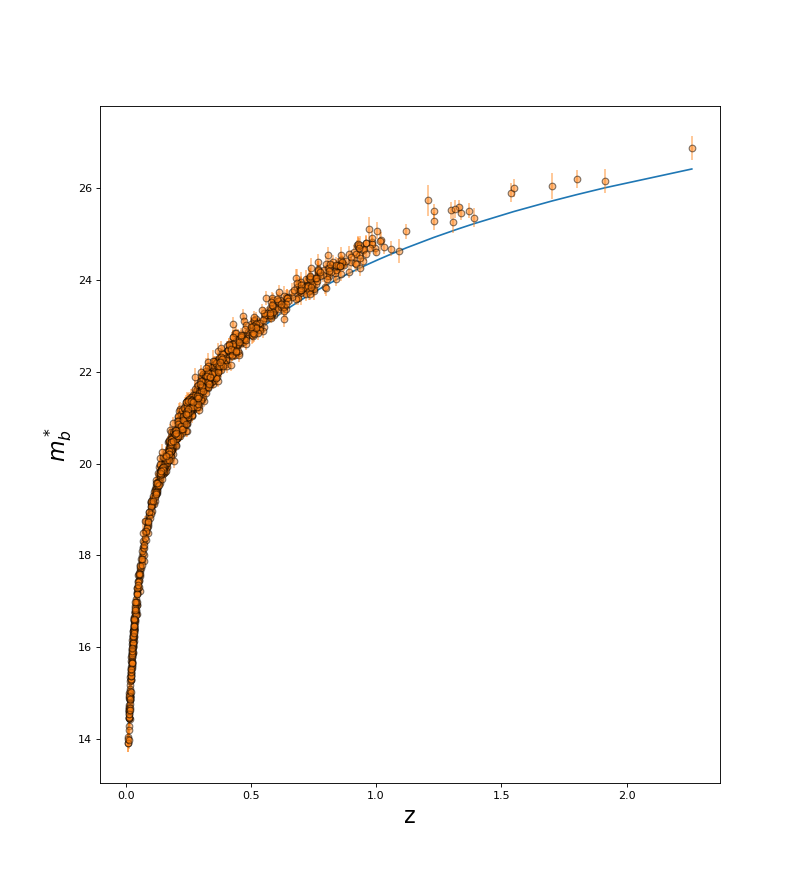}
\caption{Best fit functions for corrected magnitudes of the full Pantheon sample using the two $M(r)$ functions of the fractal smooth transition. The errors in the parameter estimation are much more lower than the sharp case, but still the fit does not perform well. }
\label{fig:6}
\end{figure}

%\begin{figure}[ht]
%\centering
%\includegraphics[scale=0.33]{FullpantALLN.png}
%\caption{Best fit functions for magnitudes of the full Pantheon sample using a fractal sharp transition, this time fitting the nuisance parameters. We get a notably worst fit than previous studies in fractal LTB models that use an inhomogeneous big bang.}
%\label{fig:5}
%\end{figure}

\subsection{Smooth transition}

The results for the best fit for the sharp transition model and details are shown in Table \ref{Table:Smooth}. We note that the errors in the parameter estimation are notably lower when the model is tested against the corrected magnitudes of the Pantheon sample, which support the idea that a smooth function performs much better in MCMC sampling. However, the $\chi^2_{red}$ values are even worse than the sharp transition model for all cases. We conclude that a flat LTB fractal model with a smooth transition, homogeneous big bang and no cosmological constant neither perform well to fit SN IA data. Best fits are plotted in Figure \ref{fig:6}.

\begin{table}[ht]
\caption{\label{Table:sharp}
Best fit values for $D$ and $L$ using a \textbf{sharp} transition, fitting for corrected magnitudes as are given in Pantheon catalogue. Also we perform a fit with a tight prior for $L$ with the corrected magnitudes to look for insights into the value of $D$.}
\begin{ruledtabular}
\textbf{Corrected magnitudes}\\
\begin{tabular}{ccccc}
 $h_{out}$ &$z_{cut}$ & $D$ & $L (Mpc)$ &$\chi^2_{red}$ \\
\hline
$0.73$ & $z<0.3$ & $3.16^{+0.36}_{-0.63}$ & $696.83^{+113.01}_{-298.50}$ & $1.15$ \\ 
$0.45$ & Full Sample & $3.47^{1.04}_{ -1.38}$ & $1354.17^{232.08}_{-313.73}$
& 1.43 \\
\end{tabular}
\newline
\newline
\textbf{Tight priors for $L$ and corrected magnitudes}\\
\begin{tabular}{cccc}
 $L$ Prior & $ D $ & $L(Mpc)$ & $\chi^2_{red}$ \\
\hline
$60h^{-1}$  & $2.97 \substack{+1.26\\-1.07}$ & 85.02 $\substack{+0.53 \\ -0.52}$  & $1.15$ \\ 
$100h^{-1}$ & $2.74 \pm \substack{+1.24\\-1.09}$ & $137.02 \substack{+0.54\\-0.51}$ & $1.15$ \\
$150h^{-1}$ & $2.73 \pm \substack{+1.2\\-1.1}$ & $205.02 \substack{+0.54\\-0.51}$ & $1.16$ \\
$200h^{-1}$ & $2.6 \pm \substack{+1.15\\-1.13}$ & $273.00 \substack{+0.51\\-0.5}$ & $1.16$ \\
\end{tabular}
\end{ruledtabular}

\end{table}

\begin{table}[ht]
\caption{\label{Table:Smooth}
Best fit values for $D$ and $L$ using a \textbf{smooth} transition, fitting for corrected magnitudes. Due to the low errors in the parameter estimation we did not extent further this approach.}
\begin{ruledtabular}
\textbf{Corrected magnitudes $M_1(r)$ function}\\
\begin{tabular}{ccccc}
 $h_{out}$ &$z_{cut}$ & $D$ & $L (Mpc)$ &$\chi^2_{red}$ \\
\hline
$0.73$ & $z<0.3$ & $3.16^{+0.018}_{-0.018}$ & $1020.86^{+0.015}_{-0.021}$ & $1.37$ \\ 
$0.45$ & Full Sample & $3.15^{0.017}_{ -0.017}$ & $1020.84^{+0.026}_{-0.014}$
& 1.46 \\
\end{tabular}
\textbf{Corrected magnitudes $M_2(r)$ function}\\
\begin{tabular}{ccccc}
 $h_{out}$ &$z_{cut}$ & $D$ & $L (Mpc)$ &$\chi^2_{red}$ \\
\hline
$0.73$ & $z<0.3$ & $3.14^{+0.019}_{-0.018}$ & $1020.86^{+0.022}_{-0.020}$ & $1.33$ \\ 
$0.45$ & Full Sample & $3.15^{+0.014}_{ -0.017}$ & $1020.85^{+0.021}_{-0.018}$
& $1.46$\\
\end{tabular}
\end{ruledtabular}

\end{table}

\section{Discussion and conclusions}
We have demonstrated that a purely fractal model following $M(r)\sim r^D$ with an homogeneous big bang is not different from an usual EdS universe with another scaling relation. As an alternative to the fractal model proposed in \cite{Cosmai19}, a more consistent model was developed, allowing for a transition between a fractal scale with $M \sim r^D$ to an homogeneous universe $M \sim r^3$ and not requiring the idea of a inhomogeneous big bang.
Two analyses to put in stress our model with the Pantheon data were performed. In the case of a sharp transition, we got a good fit for low redshift supernovae, but the model fails to describe the full behaviour of the Pantheon data sample. We conclude that fractal LTB models cannot explain the effects of dark energy. Overall, we looked for insights in fractal dimension value performing a different analysis for low redshift supernovae using tight priors for the transition scale. It was observed that the errors are still high at $z<0.3$. The results are not very different from the smooth transition case, giving a slightly worst fit than the sharp case, although the errors were much lower.
The physical validity of our models could be studied more deeply if we summarize the problems that those have. First, we did not take into account a cosmological constant. This constant could be added to the models in order to fit the full SN IA data looking for effects of the local fractal structure in the cosmological constant value. Also, we assumed that the $M$ scale should converge to a fixed $H_0$ experimental value. Two possibilities accounting for convergence to CMB $h \sim 0.45$ and another to SNIA estimation to $h \sim 0.73$ were used. Those values could be modified allowing for other definitions like presented in \cite{Camarena20} \cite{Valkenburg13}. In future works, we propose to study how a fractal structure in LTB models can impact $\Lambda$CDM cosmology. We reinforce the importance of take into account the full information about our local large scale structure to the development of a better cosmological description that lead us to stringer constraints on the $\Lambda$CDM model.
\\
\\
\\
%We conclude that is important to develop a new cosmology that accounts for the full local-large scale structure on the cosmological analysis in order to give better constrains to the dark energy problem, as is stated that it could have important effects in constraining the parameters of the $\Lambda$CDM cosmology.

\section*{ACKNOWLEDGMENTS}
EP acknowledge the support through a graduate scholarship ANID-Subdirecci\'on de Capital Humano/Doctorado Nacional/2021-21210824.

\bibliography{mybib}

\begin{thebibliography}{33}
\expandafter\ifx\csname natexlab\endcsname\relax\def\natexlab#1{#1}\fi
\expandafter\ifx\csname bibnamefont\endcsname\relax
  \def\bibnamefont#1{#1}\fi
\expandafter\ifx\csname bibfnamefont\endcsname\relax
  \def\bibfnamefont#1{#1}\fi
\expandafter\ifx\csname citenamefont\endcsname\relax
  \def\citenamefont#1{#1}\fi
\expandafter\ifx\csname url\endcsname\relax
  \def\url#1{\texttt{#1}}\fi
\expandafter\ifx\csname urlprefix\endcsname\relax\def\urlprefix{URL }\fi
\providecommand{\bibinfo}[2]{#2}
\providecommand{\eprint}[2][]{\url{#2}}

\bibitem[{\citenamefont{et~al. and Project}(1999)}]{Perlmutter99}
\bibinfo{author}{\bibfnamefont{S.~P.} \bibnamefont{et~al.}} \bibnamefont{and}
  \bibinfo{author}{\bibfnamefont{T.~S.~C.} \bibnamefont{Project}},
  \bibinfo{journal}{The Astrophysical Journal} \textbf{\bibinfo{volume}{517}},
  \bibinfo{pages}{565} (\bibinfo{year}{1999}),
  \urlprefix\url{https://doi.org/10.1086/307221}.

\bibitem[{\citenamefont{et~al.}(1998)}]{Riess98}
\bibinfo{author}{\bibfnamefont{A.~G.~R.} \bibnamefont{et~al.}},
  \bibinfo{journal}{The Astronomical Journal} \textbf{\bibinfo{volume}{116}},
  \bibinfo{pages}{1009} (\bibinfo{year}{1998}),
  \urlprefix\url{https://doi.org/10.1086/300499}.

\bibitem[{\citenamefont{Keenan et~al.}(2013)\citenamefont{Keenan, Barger, and
  Cowie}}]{Keenan13}
\bibinfo{author}{\bibfnamefont{R.~C.} \bibnamefont{Keenan}},
  \bibinfo{author}{\bibfnamefont{A.~J.} \bibnamefont{Barger}},
  \bibnamefont{and} \bibinfo{author}{\bibfnamefont{L.~L.} \bibnamefont{Cowie}},
  \bibinfo{journal}{The Astrophysical Journal} \textbf{\bibinfo{volume}{775}},
  \bibinfo{pages}{62} (\bibinfo{year}{2013}),
  \urlprefix\url{https://doi.org/10.1088/0004-637x/775/1/62}.

\bibitem[{\citenamefont{Labini}(2011)}]{Labini11}
\bibinfo{author}{\bibfnamefont{F.~S.} \bibnamefont{Labini}},
  \bibinfo{journal}{Classical and Quantum Gravity}
  \textbf{\bibinfo{volume}{28}}, \bibinfo{pages}{164003}
  (\bibinfo{year}{2011}),
  \urlprefix\url{https://doi.org/10.1088%2F0264-9381%2F28%2F16%2F164003}.

\bibitem[{\citenamefont{Labini et~al.}(1998)\citenamefont{Labini, Montuori, and
  Pietronero}}]{Labini98}
\bibinfo{author}{\bibfnamefont{F.}~\bibnamefont{Labini}},
  \bibinfo{author}{\bibfnamefont{M.}~\bibnamefont{Montuori}}, \bibnamefont{and}
  \bibinfo{author}{\bibfnamefont{L.}~\bibnamefont{Pietronero}},
  \bibinfo{journal}{Physics Reports} \textbf{\bibinfo{volume}{293}},
  \bibinfo{pages}{61} (\bibinfo{year}{1998}),
  \urlprefix\url{https://doi.org/10.1016%2Fs0370-1573%2897%2900044-6}.

\bibitem[{\citenamefont{Feindt}(2013)}]{Feindt13}
\bibinfo{author}{\bibfnamefont{U.~e.~a.} \bibnamefont{Feindt}},
  \bibinfo{journal}{A\&A} \textbf{\bibinfo{volume}{560}}, \bibinfo{pages}{A90}
  (\bibinfo{year}{2013}),
  \urlprefix\url{https://doi.org/10.1051/0004-6361/201321880}.

\bibitem[{\citenamefont{Hudson et~al.}(1999)\citenamefont{Hudson, Smith, Lucey,
  Schlegel, and Davies}}]{Hudson99}
\bibinfo{author}{\bibfnamefont{M.~J.} \bibnamefont{Hudson}},
  \bibinfo{author}{\bibfnamefont{R.~J.} \bibnamefont{Smith}},
  \bibinfo{author}{\bibfnamefont{J.~R.} \bibnamefont{Lucey}},
  \bibinfo{author}{\bibfnamefont{D.~J.} \bibnamefont{Schlegel}},
  \bibnamefont{and} \bibinfo{author}{\bibfnamefont{R.~L.}
  \bibnamefont{Davies}}, \bibinfo{journal}{The Astrophysical Journal}
  \textbf{\bibinfo{volume}{512}}, \bibinfo{pages}{L79} (\bibinfo{year}{1999}),
  \urlprefix\url{https://doi.org/10.1086%2F311883}.

\bibitem[{\citenamefont{{Magoulas} et~al.}(2016)\citenamefont{{Magoulas},
  {Springob}, {Colless}, {Mould}, {Lucey}, {Erdo{\u{g}}du}, and
  {Jones}}}]{Magoulas16}
\bibinfo{author}{\bibfnamefont{C.}~\bibnamefont{{Magoulas}}},
  \bibinfo{author}{\bibfnamefont{C.}~\bibnamefont{{Springob}}},
  \bibinfo{author}{\bibfnamefont{M.}~\bibnamefont{{Colless}}},
  \bibinfo{author}{\bibfnamefont{J.}~\bibnamefont{{Mould}}},
  \bibinfo{author}{\bibfnamefont{J.}~\bibnamefont{{Lucey}}},
  \bibinfo{author}{\bibfnamefont{P.}~\bibnamefont{{Erdo{\u{g}}du}}},
  \bibnamefont{and} \bibinfo{author}{\bibfnamefont{D.~H.}
  \bibnamefont{{Jones}}}, in \emph{\bibinfo{booktitle}{The Zeldovich Universe:
  Genesis and Growth of the Cosmic Web}}, edited by
  \bibinfo{editor}{\bibfnamefont{R.}~\bibnamefont{{van de Weygaert}}},
  \bibinfo{editor}{\bibfnamefont{S.}~\bibnamefont{{Shandarin}}},
  \bibinfo{editor}{\bibfnamefont{E.}~\bibnamefont{{Saar}}}, \bibnamefont{and}
  \bibinfo{editor}{\bibfnamefont{J.}~\bibnamefont{{Einasto}}}
  (\bibinfo{year}{2016}), vol. \bibinfo{volume}{308}, pp.
  \bibinfo{pages}{336--339}.

\bibitem[{\citenamefont{Celerier}(2006)}]{Celerier06}
\bibinfo{author}{\bibfnamefont{M.-N.} \bibnamefont{Celerier}}
  (\bibinfo{year}{2006}).

\bibitem[{\citenamefont{Enqvist}(2007)}]{Enqvist07}
\bibinfo{author}{\bibfnamefont{K.}~\bibnamefont{Enqvist}},
  \bibinfo{journal}{General Relativity and Gravitation}
  \textbf{\bibinfo{volume}{40}}, \bibinfo{pages}{451} (\bibinfo{year}{2007}),
  \urlprefix\url{https://doi.org/10.1007%2Fs10714-007-0553-9}.

\bibitem[{\citenamefont{Cosmai et~al.}(2019)\citenamefont{Cosmai, Fanizza,
  Labini, Pietronero, and Tedesco}}]{Cosmai19}
\bibinfo{author}{\bibfnamefont{L.}~\bibnamefont{Cosmai}},
  \bibinfo{author}{\bibfnamefont{G.}~\bibnamefont{Fanizza}},
  \bibinfo{author}{\bibfnamefont{F.~S.} \bibnamefont{Labini}},
  \bibinfo{author}{\bibfnamefont{L.}~\bibnamefont{Pietronero}},
  \bibnamefont{and} \bibinfo{author}{\bibfnamefont{L.}~\bibnamefont{Tedesco}},
  \bibinfo{journal}{Classical and Quantum Gravity}
  \textbf{\bibinfo{volume}{36}}, \bibinfo{pages}{045007}
  (\bibinfo{year}{2019}),
  \urlprefix\url{https://doi.org/10.1088%2F1361-6382%2Faae8f7}.

\bibitem[{\citenamefont{Tsagas}(2011)}]{Tsagas11}
\bibinfo{author}{\bibfnamefont{C.~G.} \bibnamefont{Tsagas}},
  \bibinfo{journal}{Physical Review D} \textbf{\bibinfo{volume}{84}}
  (\bibinfo{year}{2011}),
  \urlprefix\url{https://doi.org/10.1103%2Fphysrevd.84.063503}.

\bibitem[{\citenamefont{Asvesta et~al.}(2022)\citenamefont{Asvesta,
  Kazantzidis, Perivolaropoulos, and Tsagas}}]{Asvesta22}
\bibinfo{author}{\bibfnamefont{K.}~\bibnamefont{Asvesta}},
  \bibinfo{author}{\bibfnamefont{L.}~\bibnamefont{Kazantzidis}},
  \bibinfo{author}{\bibfnamefont{L.}~\bibnamefont{Perivolaropoulos}},
  \bibnamefont{and} \bibinfo{author}{\bibfnamefont{C.~G.}
  \bibnamefont{Tsagas}}, \bibinfo{journal}{Mon. Not. Roy. Astron. Soc.}
  \textbf{\bibinfo{volume}{513}}, \bibinfo{pages}{2394} (\bibinfo{year}{2022}),
  \eprint{2202.00962}.

\bibitem[{\citenamefont{Cosmai et~al.}(2023)\citenamefont{Cosmai, Fanizza,
  Labini, Pietronero, and Tedesco}}]{Cosmai2}
\bibinfo{author}{\bibfnamefont{L.}~\bibnamefont{Cosmai}},
  \bibinfo{author}{\bibfnamefont{G.}~\bibnamefont{Fanizza}},
  \bibinfo{author}{\bibfnamefont{F.~S.} \bibnamefont{Labini}},
  \bibinfo{author}{\bibfnamefont{L.}~\bibnamefont{Pietronero}},
  \bibnamefont{and} \bibinfo{author}{\bibfnamefont{L.}~\bibnamefont{Tedesco}},
  \emph{\bibinfo{title}{Comment on "a fractal ltb model cannot explain dark
  energy''}} (\bibinfo{year}{2023}),
  \urlprefix\url{https://arxiv.org/abs/2302.04679}.

\bibitem[{\citenamefont{Lemaitre}(1933)}]{Lemaitre33}
\bibinfo{author}{\bibfnamefont{G.}~\bibnamefont{Lemaitre}},
  \bibinfo{journal}{Annales Soc. Sci. Bruxelles A}
  \textbf{\bibinfo{volume}{53}}, \bibinfo{pages}{51} (\bibinfo{year}{1933}).

\bibitem[{\citenamefont{Tolman}(1934)}]{Tolman34}
\bibinfo{author}{\bibfnamefont{R.~C.} \bibnamefont{Tolman}},
  \bibinfo{journal}{Proceedings of the National Academy of Sciences}
  \textbf{\bibinfo{volume}{20}}, \bibinfo{pages}{169} (\bibinfo{year}{1934}),
  \eprint{https://www.pnas.org/doi/pdf/10.1073/pnas.20.3.169},
  \urlprefix\url{https://www.pnas.org/doi/abs/10.1073/pnas.20.3.169}.

\bibitem[{\citenamefont{Bondi}(1947)}]{Bondi47}
\bibinfo{author}{\bibfnamefont{H.}~\bibnamefont{Bondi}},
  \bibinfo{journal}{Monthly Notices of the Royal Astronomical Society}
  \textbf{\bibinfo{volume}{107}}, \bibinfo{pages}{410} (\bibinfo{year}{1947}),
  ISSN \bibinfo{issn}{0035-8711},
  \eprint{https://academic.oup.com/mnras/article-pdf/107/5-6/410/8072561/mnras107-0410.pdf},
  \urlprefix\url{https://doi.org/10.1093/mnras/107.5-6.410}.

\bibitem[{\citenamefont{Garcia-Bellido and
  Haugb{\o}lle}(2008)}]{GarciaBellido08}
\bibinfo{author}{\bibfnamefont{J.}~\bibnamefont{Garcia-Bellido}}
  \bibnamefont{and}
  \bibinfo{author}{\bibfnamefont{T.}~\bibnamefont{Haugb{\o}lle}},
  \bibinfo{journal}{Journal of Cosmology and Astroparticle Physics}
  \textbf{\bibinfo{volume}{2008}}, \bibinfo{pages}{003} (\bibinfo{year}{2008}),
  \urlprefix\url{https://doi.org/10.1088%2F1475-7516%2F2008%2F04%2F003}.

\bibitem[{\citenamefont{et~al.}(2018)}]{Scolnic18}
\bibinfo{author}{\bibfnamefont{D.~M.~S.} \bibnamefont{et~al.}},
  \bibinfo{journal}{The Astrophysical Journal} \textbf{\bibinfo{volume}{859}},
  \bibinfo{pages}{101} (\bibinfo{year}{2018}),
  \urlprefix\url{https://doi.org/10.3847/1538-4357/aab9bb}.

\bibitem[{\citenamefont{Kenworthy et~al.}(2019)\citenamefont{Kenworthy,
  Scolnic, and Riess}}]{Kenworthy19}
\bibinfo{author}{\bibfnamefont{W.~D.} \bibnamefont{Kenworthy}},
  \bibinfo{author}{\bibfnamefont{D.}~\bibnamefont{Scolnic}}, \bibnamefont{and}
  \bibinfo{author}{\bibfnamefont{A.}~\bibnamefont{Riess}},
  \bibinfo{journal}{The Astrophysical Journal} \textbf{\bibinfo{volume}{875}},
  \bibinfo{pages}{145} (\bibinfo{year}{2019}),
  \urlprefix\url{https://doi.org/10.3847%2F1538-4357%2Fab0ebf}.

\bibitem[{\citenamefont{{Hoscheit} and {Barger}}(2018)}]{Hoscheit18}
\bibinfo{author}{\bibfnamefont{B.~L.} \bibnamefont{{Hoscheit}}}
  \bibnamefont{and} \bibinfo{author}{\bibfnamefont{A.~J.}
  \bibnamefont{{Barger}}}, \bibinfo{journal}{\apj}
  \textbf{\bibinfo{volume}{854}}, \bibinfo{eid}{46} (\bibinfo{year}{2018}),
  \eprint{1801.01890}.

\bibitem[{\citenamefont{Ruffini and Stahl}(2017)}]{Ruffini17}
\bibinfo{author}{\bibfnamefont{R.}~\bibnamefont{Ruffini}} \bibnamefont{and}
  \bibinfo{author}{\bibfnamefont{C.}~\bibnamefont{Stahl}}, in
  \emph{\bibinfo{booktitle}{{14th Italian-Korean Symposium on Relativistic
  Astrophysics}}} (\bibinfo{year}{2017}).

\bibitem[{\citenamefont{Alexander et~al.}(2009)\citenamefont{Alexander, Biswas,
  Notari, and Vaid}}]{Alexander09}
\bibinfo{author}{\bibfnamefont{S.}~\bibnamefont{Alexander}},
  \bibinfo{author}{\bibfnamefont{T.}~\bibnamefont{Biswas}},
  \bibinfo{author}{\bibfnamefont{A.}~\bibnamefont{Notari}}, \bibnamefont{and}
  \bibinfo{author}{\bibfnamefont{D.}~\bibnamefont{Vaid}},
  \bibinfo{journal}{Journal of Cosmology and Astroparticle Physics}
  \textbf{\bibinfo{volume}{2009}}, \bibinfo{pages}{025} (\bibinfo{year}{2009}),
  \urlprefix\url{https://doi.org/10.1088%2F1475-7516%2F2009%2F09%2F025}.

\bibitem[{\citenamefont{Calcagni}(2017)}]{Calcagni_2017}
\bibinfo{author}{\bibfnamefont{G.}~\bibnamefont{Calcagni}},
  \bibinfo{journal}{Journal of High Energy Physics}
  \textbf{\bibinfo{volume}{2017}} (\bibinfo{year}{2017}),
  \urlprefix\url{https://doi.org/10.1007%2Fjhep03%282017%29138}.

\bibitem[{\citenamefont{Dainotti et~al.}(2021)\citenamefont{Dainotti, Simone,
  Schiavone, Montani, Rinaldi, and Lambiase}}]{Dainotti_2021}
\bibinfo{author}{\bibfnamefont{M.~G.} \bibnamefont{Dainotti}},
  \bibinfo{author}{\bibfnamefont{B.~D.} \bibnamefont{Simone}},
  \bibinfo{author}{\bibfnamefont{T.}~\bibnamefont{Schiavone}},
  \bibinfo{author}{\bibfnamefont{G.}~\bibnamefont{Montani}},
  \bibinfo{author}{\bibfnamefont{E.}~\bibnamefont{Rinaldi}}, \bibnamefont{and}
  \bibinfo{author}{\bibfnamefont{G.}~\bibnamefont{Lambiase}},
  \bibinfo{journal}{The Astrophysical Journal} \textbf{\bibinfo{volume}{912}},
  \bibinfo{pages}{150} (\bibinfo{year}{2021}),
  \urlprefix\url{https://doi.org/10.3847%2F1538-4357%2Fabeb73}.

\bibitem[{\citenamefont{Alnes et~al.}(2005)\citenamefont{Alnes, Amarzguioui,
  and Gron}}]{Alnes}
\bibinfo{author}{\bibfnamefont{H.}~\bibnamefont{Alnes}},
  \bibinfo{author}{\bibfnamefont{M.}~\bibnamefont{Amarzguioui}},
  \bibnamefont{and} \bibinfo{author}{\bibfnamefont{O.}~\bibnamefont{Gron}},
  \bibinfo{journal}{Physical Review D} \textbf{\bibinfo{volume}{73}}
  (\bibinfo{year}{2005}).

\bibitem[{\citenamefont{{Foreman-Mackey}
  et~al.}(2013)\citenamefont{{Foreman-Mackey}, {Hogg}, {Lang}, and
  {Goodman}}}]{2013PASP..125..306F}
\bibinfo{author}{\bibfnamefont{D.}~\bibnamefont{{Foreman-Mackey}}},
  \bibinfo{author}{\bibfnamefont{D.~W.} \bibnamefont{{Hogg}}},
  \bibinfo{author}{\bibfnamefont{D.}~\bibnamefont{{Lang}}}, \bibnamefont{and}
  \bibinfo{author}{\bibfnamefont{J.}~\bibnamefont{{Goodman}}},
  \bibinfo{journal}{\pasp} \textbf{\bibinfo{volume}{125}}, \bibinfo{pages}{306}
  (\bibinfo{year}{2013}), \eprint{1202.3665}.

\bibitem[{\citenamefont{{Goodman} and {Weare}}(2010)}]{2010CAMCS...5...65G}
\bibinfo{author}{\bibfnamefont{J.}~\bibnamefont{{Goodman}}} \bibnamefont{and}
  \bibinfo{author}{\bibfnamefont{J.}~\bibnamefont{{Weare}}},
  \bibinfo{journal}{Communications in Applied Mathematics and Computational
  Science} \textbf{\bibinfo{volume}{5}}, \bibinfo{pages}{65}
  (\bibinfo{year}{2010}).

\bibitem[{\citenamefont{{Yadav} et~al.}(2010)\citenamefont{{Yadav}, {Bagla},
  and {Khandai}}}]{2010MNRAS.405.2009Y}
\bibinfo{author}{\bibfnamefont{J.~K.} \bibnamefont{{Yadav}}},
  \bibinfo{author}{\bibfnamefont{J.~S.} \bibnamefont{{Bagla}}},
  \bibnamefont{and}
  \bibinfo{author}{\bibfnamefont{N.}~\bibnamefont{{Khandai}}},
  \bibinfo{journal}{\mnras} \textbf{\bibinfo{volume}{405}},
  \bibinfo{pages}{2009} (\bibinfo{year}{2010}), \eprint{1001.0617}.

\bibitem[{\citenamefont{{Yadav} et~al.}(2005)\citenamefont{{Yadav},
  {Bharadwaj}, {Pandey}, and {Seshadri}}}]{2005MNRAS.364..601Y}
\bibinfo{author}{\bibfnamefont{J.}~\bibnamefont{{Yadav}}},
  \bibinfo{author}{\bibfnamefont{S.}~\bibnamefont{{Bharadwaj}}},
  \bibinfo{author}{\bibfnamefont{B.}~\bibnamefont{{Pandey}}}, \bibnamefont{and}
  \bibinfo{author}{\bibfnamefont{T.~R.} \bibnamefont{{Seshadri}}},
  \bibinfo{journal}{\mnras} \textbf{\bibinfo{volume}{364}},
  \bibinfo{pages}{601} (\bibinfo{year}{2005}), \eprint{astro-ph/0504315}.

\bibitem[{\citenamefont{{Hogg} et~al.}(2005)\citenamefont{{Hogg}, {Eisenstein},
  {Blanton}, {Bahcall}, {Brinkmann}, {Gunn}, and
  {Schneider}}}]{2005ApJ...624...54H}
\bibinfo{author}{\bibfnamefont{D.~W.} \bibnamefont{{Hogg}}},
  \bibinfo{author}{\bibfnamefont{D.~J.} \bibnamefont{{Eisenstein}}},
  \bibinfo{author}{\bibfnamefont{M.~R.} \bibnamefont{{Blanton}}},
  \bibinfo{author}{\bibfnamefont{N.~A.} \bibnamefont{{Bahcall}}},
  \bibinfo{author}{\bibfnamefont{J.}~\bibnamefont{{Brinkmann}}},
  \bibinfo{author}{\bibfnamefont{J.~E.} \bibnamefont{{Gunn}}},
  \bibnamefont{and} \bibinfo{author}{\bibfnamefont{D.~P.}
  \bibnamefont{{Schneider}}}, \bibinfo{journal}{\apj}
  \textbf{\bibinfo{volume}{624}}, \bibinfo{pages}{54} (\bibinfo{year}{2005}),
  \eprint{astro-ph/0411197}.

\bibitem[{\citenamefont{Camarena and Marra}(2020)}]{Camarena20}
\bibinfo{author}{\bibfnamefont{D.}~\bibnamefont{Camarena}} \bibnamefont{and}
  \bibinfo{author}{\bibfnamefont{V.}~\bibnamefont{Marra}},
  \bibinfo{journal}{Physical Review Research} \textbf{\bibinfo{volume}{2}}
  (\bibinfo{year}{2020}),
  \urlprefix\url{https://doi.org/10.1103%2Fphysrevresearch.2.013028}.

\bibitem[{\citenamefont{Valkenburg et~al.}(2013)\citenamefont{Valkenburg,
  Marra, and Clarkson}}]{Valkenburg13}
\bibinfo{author}{\bibfnamefont{W.}~\bibnamefont{Valkenburg}},
  \bibinfo{author}{\bibfnamefont{V.}~\bibnamefont{Marra}}, \bibnamefont{and}
  \bibinfo{author}{\bibfnamefont{C.}~\bibnamefont{Clarkson}},
  \bibinfo{journal}{Monthly Notices of the Royal Astronomical Society: Letters}
  \textbf{\bibinfo{volume}{438}}, \bibinfo{pages}{L6} (\bibinfo{year}{2013}),
  \urlprefix\url{https://doi.org/10.1093%2Fmnrasl%2Fslt140}.

\end{thebibliography}

\end{document}